\documentclass[twocolumn]{article}

\onecolumn

\title{\sf Application Software,\\
           Domain-Specific Languages,\\ and\\
           Language Design Assistants}

\author {{\sf Jan Heering}\\
         {\sf\em\footnotesize CWI}\\
         {\sf\em\footnotesize P.O. Box 94079, 1090 GB Amsterdam, The Netherlands}\\
         {\tt\footnotesize Jan.Heering@cwi.nl}}
     
\date{}

\begin{document}
\maketitle

\begin{quote}
\footnotesize
        \noindent {\sf 
        ABSTRACT\\
While application software does the real work, domain-specific
languages (DSLs) are tools to help produce it efficiently, and
language design assistants in turn are meta-tools to help produce DSLs
quickly.  DSLs are already in wide use (HTML for web pages, Excel
macros for spreadsheet applications, VHDL for hardware design,
$\ldots$), but many more will be needed for both new as well as
existing application domains. Language design assistants to help
develop them currently exist only in the basic form of language
development systems.  After a quick look at domain-specific languages,
and especially their relationship to application libraries, we survey
existing language development systems and give an outline of future
language design assistants.
}

        \noindent {\sf 
        {\em 1991 Computing Reviews Classification System:} D.3}

        \noindent {\sf 
        {\em Keywords and Phrases:} 
        application software, application domain, domain-specific language,
        task-oriented language, end user programming, language development system,
        language-based tool generation, language design assistant, language concept,
        description by example.}

        \noindent {\sf 
        {\em Note:} To be presented at SSGRR 2000, L'Aquila, Italy.
        This research was supported in part by
        the Telematica Instituut under the {\em Domain-Specific Languages} project.}

\end{quote}

\twocolumn

\section{Domain-Specific Languages} \label{sec:DSLs}

Many computer languages are \textit{domain-specific}
rather than general purpose. Domain-specific languages (DSLs) are also
called \textit{task-specific}, \textit{application-oriented}, or
\textit{problem-oriented}.  Some well-known DSLs with their
application domains are listed in Table~\ref{table:DSLexamples}.
So-called \textit{fourth-generation languages} (4GLs) are usually DSLs
for database applications.  A good reference, which focuses on the key role
of DSLs in end user programming, is \cite{Nardi93}.  A recent DSL
bibliography is \cite{DeursenEtAl00}.

We will not try to give a definition of what constitutes an
application domain and what does not.  Some people consider Cobol to
be a DSL for business applications, while others would argue this is
pushing the notion of application domain too far.  Leaving matters of
definition aside, it is natural to think of DSLs in terms of a gradual
scale with very specialized DSLs such as HTML
(Table~\ref{table:DSLexamples}) on the left and general purpose
programming languages such as C++ on the right. On this scale, Cobol
would be somewhere between HTML and C++, but much closer to the
latter.

In combination with an \textit{application library}, any general purpose
programming language can act as a DSL, so why were DSLs developed in the
first place? Simply because they can offer domain-specificity in
better ways:
\begin{itemize}

\item Appropriate or established \textit{domain-specific notations} 
      are usually beyond the limited user-definable operator notation
      offered by general purpose languages. A DSL offers
      domain-specific notations from the start. Their importance
      cannot be overestimated as they are directly related to
      the suitability for end user programming and, more generally,
      the programmer productivity improvement associated with the
      use of DSLs.

\item Appropriate \textit{domain-specific constructs and abstractions} cannot
      always be mapped in a straightforward way on functions or
      objects that can be put in a library.  This means a general
      purpose language using an application library can only express
      these constructs indirectly. Again, a DSL would incorporate
      domain-specific constructs from the start.

\end{itemize}
Nevertheless, application libraries are formidable competitors to
DSLs. For once, designing and implementing a DSL is far from easy.  A
domain expert is usually not an expert in language design, which is a
distinct (meta-)domain of expertise in itself.  How this situation can be
improved is  the main topic of this article, but even with improved DSL
development tools, application libraries will remain the most
cost-effective solution in many cases.

\begin{table}
\begin{center}
\caption{Some widely used domain-specific languages.} \label{table:DSLexamples}
\begin{tabular}{|p{4cm}|p{4cm}|}
\hline
DSL  & Application domain \\ \hline \hline

BNF  & Syntax \\

Excel macro language & Spreadsheets \\

HTML & Hypertext web pages \\

\LaTeX & Typesetting \\

Make & Program maintenance \\

SQL  & Database queries \\

VHDL & Hardware design \\

\hline

\end{tabular}
\end{center}
\end{table}

There are other factors complicating the relative merits of DSLs and
application libraries. A case in point is Microsoft Excel. Its macro
language is a DSL for spreadsheet applications which adds
programmability to Excel's fundamental interactive mode. Using COM,
Microsoft's software component technology, Excel's implementation has
been restructured into an application library or tool box of COM
components.  This has opened it up to general purpose programming
languages such as C++, Java and Basic, which can access it through its
COM interfaces.  This is called \textit{Automation} and is described
in more detail in
\cite{Chappell96}.  Unlike Excel macro language, which by its very
nature is limited to Excel functionality, general purpose programming
languages are not. They can be used to write applications transcending
Excel's boundaries by using components from other ``automated'' programs
and COM libraries in addition to components from Excel itself.

\section{Existing Language Development Systems} \label{sec:SYSTEMS}

As noted, DSL development is hard, requiring both domain knowledge and
language development expertise. The development process can be speeded
up by using a \textit{language development system}. Some
representative ones are listed in Table~\ref{table:SYSTEMS}.  They
have widely different capabilities and are in widely different stages
of development, but are based on the same general principle:
\textit{they generate programming tools from language descriptions.}

The input to these systems is a description of various aspects of the
DSL to be developed in terms of specialized meta-languages.  
Some important language aspects are listed in Table~\ref{table:ASPECTS}.
It so happens that the meta-languages used for describing these aspects are themselves DSLs for the
particular aspect in question.  For instance, an important
language aspect is syntax, which is usually described in something
close to BNF, the well-known DSL and \textit{de facto} standard for syntax specification
(Table~\ref{table:DSLexamples}).  The corresponding tool generated by
the language development system is a parser.

The tool generation capabilities of the language development systems
listed in Table~\ref{table:SYSTEMS} are shown in
Table~\ref{table:CAPABILITIES}. All of them can generate lexical
scanners, parsers, and prettyprinters, many of them can produce
syntax-directed editors, typecheckers, and interpreters, and a few can
produce various kinds of software renovation tools. These tools are as
useful for DSLs as they are for programming languages.
 
Although the various specialized meta-languages used
for describing language aspects differ from system to system, they are usually
\textit{rule based}. For instance, depending on the system, the typechecking of
language constructs has to be described in terms of
\textit{attributed syntax rules} (an extension of BNF),
\textit{conditional rewrite rules}, \textit{inference rules}, or
\textit{transition rules}.

Some examples of DSL development using a language development system
are given in Table~\ref{table:DSLdevelopment}.  The Box prettyprinting
meta-language is an example of a DSL developed with a language
development system (in this case the ASF+SDF Meta-En\-vi\-ron\-ment)
for later use as one of the meta-languages of the system itself. Risla
is a language for describing loans and mortgages offered by banks. Its
constructs are described in terms of their translation to Cobol. The
LaCon system is not listed among the language development systems in
Table~\ref{table:SYSTEMS}, but will be discussed in the next
section.

\begin{table}
\begin{center}
\caption{Some representative language development systems.} \label{table:SYSTEMS}
\begin{tabular}{|p{4cm}|p{4cm}|}
\hline
System & Developed at    \\ 
\hline \hline

ASF+SDF Meta-En\-vi\-ron\-ment \cite{DHK96} & CWI and University of Amsterdam \\

Centaur \cite{BorrasEtAl89} & INRIA Sophia-Anti\-po\-lis \\

Eli \cite{KastensEtAl98} & University of Paderborn \\
     
Gem-Mex \cite{AnlauffEtAl97} & University of L'Aquila \\

PSG \cite{BahlkeSnelting86} & Technical University of Darmstadt \\
   
Software Refinery \cite{MarkosianEtAl94} & Reasoning Systems, Palo Alto \\

Synthesizer Generator \cite{RT89} & Cornell University \\

\hline
\end{tabular}
\end{center}
\end{table}

\begin{table}
\begin{center}
\caption{Some language aspects.} \label{table:ASPECTS}
\begin{tabular}{|c|}
\hline

Syntax \\
Prettyprinting \\
Typechecking \\
Interpretation \\
Translation \\
Debugging \\

\hline
\end{tabular}
\end{center}
\end{table}

\begin{table*}[p]
\begin{center}
\caption{Tool generation capabilities of representative language development systems.} \label{table:CAPABILITIES}
\begin{tabular}{|p{5cm}|p{10cm}|}
\hline
System                         & Generated tools \\ \hline \hline

ASF+SDF Meta-En\-vi\-ron\-ment & Scanner/parser (generalized LR), 
                                 prettyprinter, syntax-directed editor,
                                 typechecker, interpreter,
                                 origin tracker, translator,
                                 renovation tools, $\ldots$
\\ \hline

Centaur                        & Scanner/parser (LALR),
                                 prettyprinter, syntax-directed editor,
                                 typechecker, interpreter,
                                 origin tracker, translator, $\ldots$ 
\\ \hline

Eli                            & Scanner/parser, typechecker, interpreter, translator, $\ldots$
\\ \hline

Gem-Mex                        & Scanner/parser, typechecker, interpreter, translator,
                                 debugger, $\ldots$ 
\\ \hline

PSG                            & Scanner/parser, syntax-directed editor,
                                 incremental typechecker (even for incomplete
                                 program fragments), interpreter                 
\\ \hline
        
Software Refinery              & Scanner/parser (LALR), prettyprinter,
                                 syntax-directed editor, object-oriented
                                 parse tree repository (including dataflow
                                 relations), 
                                 Y2K/Euro tools, program slicer, $\ldots$
\\ \hline

Synthesizer Generator          & Scanner/parser (LALR), prettyprinter,
                                 syntax-directed editor, incremental typechecker,
                                 incremental translator, $\ldots$
\\ \hline

\end{tabular}
\end{center}
\end{table*}

\begin{table*}[p]
\begin{center}
\caption{Examples of DSL development using a language development system.} \label{table:DSLdevelopment}
\begin{tabular}{|p{4cm}|p{4cm}|p{4cm}|p{4cm}|}
\hline
DSL  & Application domain & System used \\ \hline \hline

Box \cite{BV96} & Prettyprinting & ASF+SDF Meta-En\-vi\-ron\-ment \\

Cubix \cite{KutterEtAl98} & Virtual data warehousing & Gem-Mex \\

Risla \cite{DK98} & Financial products & ASF+SDF Meta-En\-vi\-ron\-ment \\

(Various) & Data model translation & LaCon \cite{KastensPfahler98} \\

\hline
\end{tabular}
\end{center}
\end{table*}

\begin{table*}[p]
\begin{center}
\caption{Some language concepts.} \label{table:CONCEPTS}
\begin{tabular}{|p{2cm}|p{2cm}|p{2cm}|p{2cm}|p{2cm}|p{2cm}|p{2cm}|}
\hline
syntax  & sentence  & abstract syntax & signature & term & &
\\ \hline \hline
program & procedure & function & statement & expression & assignment & goto 
\\ \hline      
conditional & loop & exception handler & input & output & operand & value  
\\ \hline \hline
type & subtype & inheritance & strong typing & overloading & polymorphism & untyped 
\\ \hline \hline
record & class & subclass & template class & object & module & parameterized module
\\ \hline \hline
import & export & actualization & instantiation & external & library &
\\ \hline \hline
concurrency & thread & process & exception & overflow & &
\\ \hline \hline
scope & block & static scope & dynamic scope & local variable & global variable & static variable 
\\ \hline
file & parameter & argument & & & &
\\ \hline \hline
state & heap & call stack & stack frame & side-effect & data flow & control flow 
\\ \hline \hline
typechecking & interpretation & translation & transforma\-tion & abstract interpretation & data flow analysis & control flow analysis 
\\ \hline 
\end{tabular}
\end{center}
\end{table*}

\section{Toward Language Design Assistants} \label{sec:LDA}

Actually, the language development systems listed in
Table~\ref{table:SYSTEMS} incorporate little language design
knowledge. Their main assets are the meta-languages they support, and
in some cases a meta-environment to aid in constructing and debugging
language descriptions.  Even though tailored toward the language
aspect they have to describe, these meta-languages are often hard
to use. In many cases they were selected primarily for their favorable
mathematical or logical properties rather than their user-friendliness.
Also, these properties tend to become less important when
language descriptions become large.

To turn language development systems into true \textit{language design assistants}
(LDAs) \cite{HeeringKlint00}, improvements can be sought in various directions:

\begin{itemize}

\item Incorporation of \textit{language concepts} and \textit{design rules}.

\item \textit{Visual} or \textit{semi-visual} meta-languages. The latter
      are partly graphical and partly textual.

\item \textit{Description by example} of some (necessarily limited) language aspects
      such as prettyprinting or syntax.

\end{itemize}

Clearly, these improvements are not simple. LDAs will become far more
complex than language development systems, which are not particularly
simple to begin with. Fortunately, LDAs may be approached step by
step. Also, not all of the above improvements are completely new:

\begin{itemize}
 
\item The LaCon system
(Table~\ref{table:DSLdevelopment}) allows domain experts to compose
elements and properties of a DSL by simple yes/no decisions.  It
automatically checks consistency of user decisions, computes their
consequences, and provides design style advice.  To generate an
implementation, it uses the Eli language development system
(Table~\ref{table:SYSTEMS}) as back-end.

\item The Gem-Mex system (Table~\ref{table:SYSTEMS}) already supports a
semi-visual notation for the transition rules it uses to define the
typechecking, interpretation and translation of language constructs.

\item Less concretely, no longer pursued but nevertheless interesting
plans for the Language Development Laboratory included a library of
reusable language constructs, a knowledge base containing knowledge of
languages and their compilers/interpreters, and a tool for language
design \cite{HarmEtAl97,Laemmel97}.

\end{itemize}

In the remainder of this section we focus on the
incorporation of language design knowledge and on description by
example of selected language aspects.
  
\subsection{Incorporation of Language Concepts and Design Rules}

Basically, the language designer using an LDA picks suitable language
building blocks from the language knowledge base (language library),
customizes them, and composes them into larger and larger language
fragments. It may be necessary to add entirely new building blocks and
concepts in the process, especially domain-specific ones, since it
cannot be expected that everything required is already present in
customizable form. The LDA provides feedback during customization
and composition. Finally, the finished design is implemented by a
language development system that serves as back-end to the LDA.

More specifically, the main elements and notions that would play a key role
in an LDA are:

\begin{itemize}

\item \textbf{Language concept} 
      Some examples are given in Table~\ref{table:CONCEPTS}.  They are
      rather diverse. For instance, some language concepts, such as
      ``statement'', ``expression'', or ``loop'', correspond more or
      less directly to language constructs. These are
      \textit{language building blocks} (see below).  Other ones, such
      as ``scope'' or ``side-effect'' have the character of an
      attribute to a language building block, while ``state'' or
      ``call stack'' refer to the dynamic behavior of programs or to
      the language's implementation. Note that some of the concepts in
      the bottom row correspond to language aspects mentioned in the
      previous section. 
      
      A more precise classification of language
      concepts from the perspective of their use in an LDA is
      currently being attempted. Some concepts may turn out to be
      more useful than others.

\item \textbf{Language building block} \textit{Language concept}
      corresponding more or less directly to a language construct,
      such as ``statement'', ``expression'', or ``loop''.  It may have
      many attributes, which themselves correspond to \textit{language
      concepts} of a different kind, such as (abstract) syntax,
      typechecking, interpretation, data and control flow,
      side-effects, exceptions, among others.
   
\item \textbf{Relation} 
      Relation between \textit{language concepts}. May itself be a
      (higher-order) \textit{language concept}. Attributes like
      ``scope'' and ``side-effect'' are unary relations.
      ``Implementation'' would be an example of a binary one.

\item \textbf{Language knowledge base} Knowledge base of
      \textit{language concepts} and their \textit{relations}.
      It may include theories of the  concepts it contains. 
      These may range from rudimentary to elaborate.

\item \textbf{Customization} The process of adapting  
      \textit{language building blocks} during the language design
      process by means of instantiation, transformation, or
      generation.

\item \textbf{Composition} (Partly) \textit{customized}
      \textit{language building blocks} can be composed into larger
      language fragments. In this way the DSL is constructed step by
      step.

\item \textbf{Constraint checking} is triggered by \textit{customization}
      and \textit{composition} to check, at least to some extent, the
      validity of the resulting \textit{language building block}. The
      constraint checker is a parameter of the LDA design.

\item \textbf{Design language} Meta-language allowing the user to formulate language
      design questions and interrogate/browse the \textit{language
      knowledge base}. 

\item \textbf{Language knowledge representation language} 
      Visual or semi-visual meta-language to express \textit{language
      concepts} and their \textit{relations}.  It has well-defined
      sublanguages to express \textit{language building blocks} and
      their attributes. These sublanguages are compiled to the
      corresponding meta-language(s) of the \textit{language
      development system} that is used as a back-end.

\item \textbf{Language development system} Back-end
      of the LDA and parameter of the LDA design. The LDA facilitates
      the language description process and then uses a language
      development system to generate the tooling from the finished
      description.  As noted, LaCon uses Eli as back-end.  In our
      case, the ASF+SDF Meta-Environment (Table~\ref{table:SYSTEMS})
      will be the back-end. Many of the other systems mentioned in the
      previous section would be suitable as well.

\end{itemize}

\subsection{Description by Example of Selected Language Aspects}

For some language aspects description by example (DBE) may become a
user-friendly alternative to (or addition to) exhaustive description in
terms of a specialized meta-language.  We should not be too
ambitious. DBE of the interpretation or translation of a language is
not realistic, but prettyprinting or syntax may be sufficiently limited
for DBE to become useful.

In fact, prettyprinting is a special case of text formatting.  DBE of
the latter is supported to some extent by Microsoft Word, which is
capable of defining ``styles'' from user supplied examples.  Another
system supporting formatting by example is Tourmaline
\cite{Myers93}. It contains rules that try to determine
the role of different parts of a header in a text document, such as
section number, title, author, and affiliation, as well as the formatting
associated with each part. The results are displayed in a dialogue box
for the user to inspect and correct. 

DBE of syntax for the purpose of language development was already
suggested in \cite{CrespiEtAl73}, but to the best of our knowledge not
put into practice.  Since then, considerable progress has been made
both in the theory of syntax inference from example sentences as well
as in the computing power that can be brought to bear on the inference
process.

The user-friendliness of DBE is due to the fact that examples of
intended behavior do not require a specialized meta-language, or only
a small part of it. For instance, prettyprinting by example would try
to infer general prettyprinting rules from a test suite of
prettyprinted constructs or programs in the language under
development.  The inferred rules might be expressed in a language like
Box (Table~\ref{table:DSLdevelopment}), the prettyprinting
meta-language of the ASF+SDF Meta-En\-vi\-ron\-ment, but the test
suite itself would not require this language.

The inference mechanism used in the implementation of DBE may range
from very limited and predictable generalization, hardly deserving to
be called inference, to full-fledged inductive inference.  The latter
has a probabilistic character in the sense that the inferred
generalization is in some sense the simplest one that is correct on
the examples given to it, but it need not be the one intended by the
user. To find out, the user may have to inspect the generated rules,
which is rather unattractive. And if the generalization turns
out to be incorrect, it may again be hard to find out which examples have to
be added or changed.

These are largely unsolved problems, partly offsetting the advantages
of the stronger forms of DBE suffering from them, so simple,
predictable DBE is to be preferred.

\section*{Acknowledgement}
 
The basic idea for a language design assistant emerged in discussions
with Paul Klint.

\bibliographystyle{abbrv}

\end{document}